# Structured Functional Principal Component Analysis

Haochang Shou[a], Vadim Zipunnikov[a], Ciprian M. Crainiceanu[a] and Sonja Greven[b]

April 24, 2013

## Abstract

Motivated by modern observational studies, we introduce a class of functional models that expands nested and crossed designs. These models account for the natural inheritance of correlation structure from sampling design in studies where the fundamental sampling unit is a function or image. Inference is based on functional quadratics and their relationship with the underlying covariance structure of the latent processes. A computationally fast and scalable estimation procedure is developed for ultra-high dimensional data. Methods are illustrated in three examples: high-frequency accelerometer data for daily activity, pitch linguistic data for phonetic analysis, and EEG data for studying electrical brain activity during sleep.

[a]Department of Biostatistics, Johns Hopkins University, 615 N. Wolfe St., Baltimore, MD 21205, USA
[b]Department of Statistics, Ludwig-Maximilians-Universität München, 80539 München, Germany
Corresponding Author: Haochang Shou, Phone: 410-614-5086. Email: hshou@jhsph.edu.

# 1 Introduction

In many current studies, functional measurements have well defined stochastic structure induced either by the experimental design or by the scientific meaning of the data. For example, the Sleep Heart Health Study (SHHS) [Quan *et al.* (1997); Crainiceanu *et al.* (2009); Di *et al.* (2009)] collected electroencephalograms (EEG) data for thousands of subjects at two visits, roughly five years apart. At every visit, EEG data were recorded at 125Hz during sleep. Thus, for each subject and visit, data consist of 125 observations per second, or 0.45 million per hour. Crainiceanu *et al.* (2009) applied a Fourier transformation to the original data and obtained the normalized $\delta$-power as a densely-sampled stationary time series. These data have a natural hierarchical structure induced by the replicated visits within each subject. To be more precise, one can denote the $\delta$-power function for visit $j$ of subject $i$ at time $t$ after sleep onset by $Y_{ij}(t)$. $Y_{ij}(t)$ can be decomposed into a subject-specific process $X_i(t)$ and a visit-within-subject process $U_{ij}(t)$ that quantifies the deviation from the subject-specific mean.

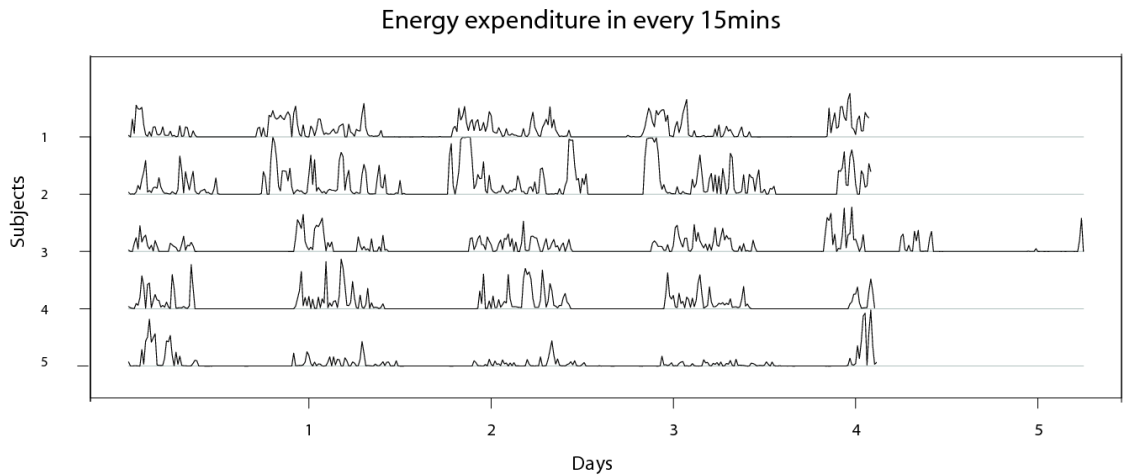

**Figure 1:** *Activity intensity measurements over 5 days for 4 subjects. Original data has 10 observations per second. This plot shows the average of activity intensity in non-overlapping 15 minute intervals for improved display clarity.*

A second example is provided by Bai *et al.* (2012) in a recent study of physical activity in an elderly population. In this study, each subject wears an accelerometer that records three-axis accelerations during in-home activities at a sampling frequency of 10Hz. After normalization with respect to at-rest variability, Bai *et al.* (2012) introduced activity intensity, a measure of activity expressed in multiple of signal standard deviations during inactivity periods. Activity intensity is calculated in every tenth of one second interval. Figure 1 displays the activity intensity for five subjects during a 5-day period averaged over 15 minutes for improved display clarity. One possibility of analyzing these data is to focus on activity intensity in non-overlapping one-hour intervals. Thus, the data for each subject contains five days, each with $36,000$ activity measurements per hour for 24 hours. This can be viewed as a three-level hierarchical structure: hour within day within subject. More specifically, let $Y_{ijk}(t)$ be the activity intensity at time $t$ within hour $k$ on day $j$ for subject $i$. In addition to the subject-specific process $X_i(t)$ and the day-within-subject process $U_{ij}(t)$, the remaining part of the variation in $Y_{ijk}(t)$ can be explained by the hour-specific process $W_{ijk}(t)$ that quantifies the deviation of hour $k$ from the average of day $j$ for subject $i$.

Aston *et al.* (2010) described a different study of phonetic analysis where the authors were interested in studying the fundamental frequency (F0,'pitch') of spoken language. In particular, they recorded the F0-contours of syllables from 19 nouns pronounced by 8 native speakers of the Luobuzhai Qiang dialect in China. Suppose that we use $Y_{ij}(t)$ to denote the pitch of syllables $j$ at normalized vowel time $t$ spoken by subject $i$. Each $Y_{ij}(t)$ was measured at 11 equidistant time points across the duration of a vowel. The value of $Y_{ij}(t)$ is jointly affected by at least two components: the syllable-generic effect $X_i(t)$ relating to the word, tone, stress and intonation of the syllable, and the speaker-inherent effect $Z_j(t)$ that could depend on age, gender etc. Unlike the hierarchical framework in the previous examples, these two latent processes can be assumed to be mutually independent while they can have interactive effect on the pitch contours. Figure 2 displays an example of F0-contours for vowels that compose three different words spoken by three speakers. The color stratification indicates that some speakers have, on average, a lower pitch than others. The shape of the curves are strongly correlated with the vowels and words that they belong to. For instance, within vowel 'i', curves from word 3 all display a steep rising pattern and go down at the end of the vowel. But curves from word 2 (labeled by the triangle symbol) are all arch-shaped.

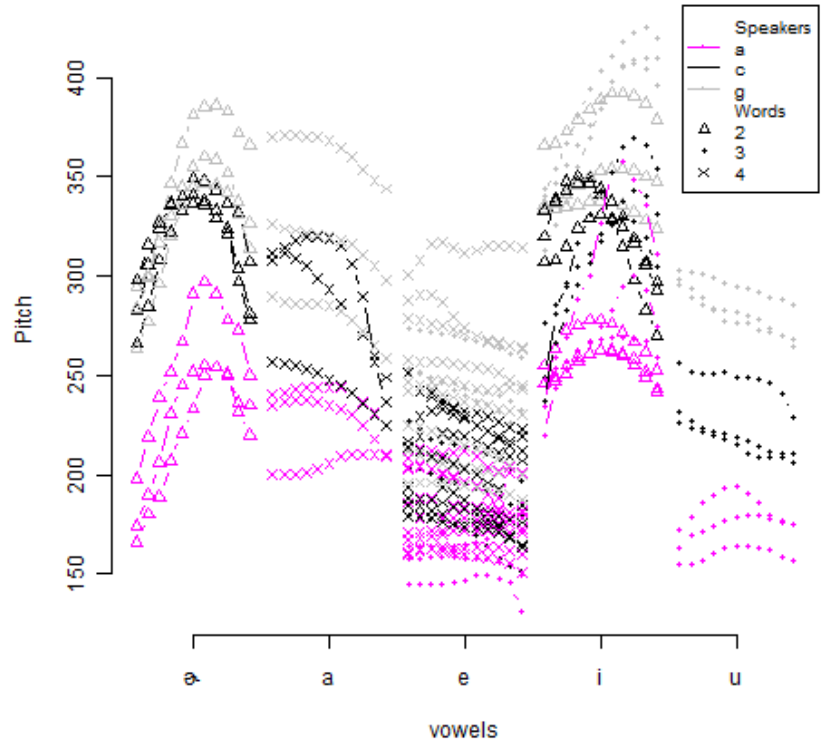

**Figure 2:** *F0-contours for 3 words (triangles for word 2, dots for word 3, and multiplication signs for word 4) spoken by 3 speakers (pink for speaker 'a', black for speaker 'c', and gray for speaker 'g'). Each contour was measured at 11 equal distant time points within a vowel ('ə','a','e','i' or 'u') when a particular word was spoken by one of the eight speakers. Every word was repeated under three different contexts.*

Although these three studies have different designs, they share some common features: 1) the fundamental observation unit is a function that can be very high-dimensional; 2) data has a known structure induced by the sampling design; and 3) analysis of individual levels of variability is of interest. One goal of this paper is to define a wide class of structured functional models with explicit functional effect components; in particular, this class will contain the observed structures in the three examples. We will focus on the common structures and provide a consistent statistical framework for all these models. A second goal is to characterize the observed variability by uncorrelated latent processes, whose covariance estimators are obtained using method of moments estimators from the observed data. Through estimating and diagonalization of these covariance operators, we will achieve both dimensionality reduction of the original data and statistical modeling on the induced linear spaces. From an intuitive perspective, this paper shows how to conduct statistically principled PCA when the data matrix has a particular known and common latent correlation structure.

The structured functional models in this paper fall into functional linear mixed model (FLMM) framework. Earlier work [Guo (2002); Herrick & Morris (2006); Morris & Carroll (2006)] mainly used splines or wavelets smoothing in model fitting. Brumback & Rice (1998) and Guo (2004) have specifically studied functional nested and crossed designs. More recent work like Staicu *et al.* (2010) and Zhou *et al.* (2010) considered spatial correlation in the nested model. While all these models could be viewed as particular cases of FLMM, model fit and inference remains difficult and it is currently done on a model-by-model basis. We conclude that none of these previous papers have addressed the class of complex functional structures we discuss here. Moreover, we are the first to propose very fast approaches for high-dimensional data. We aim to introduce a data-driven approach that applies to both nested and crossed designs, but is generalizable to a much broader model space. We achieve this by introducing latent processes that capture explicit levels of variability using the same concept from standard mixed effects models. The only difference is that random effects are now replaced with random processes. Computational feasibility is achieved via principal component decomposition of latent processes covariance operators and by lossless projections for ultra-high dimensional data. These approaches are methodologically related to the PCA decomposition. [Staniswalis & Lee (1998); Yao *et al.* (2003); Yao *et al.* (2005); Di *et al.* (2009) (MFPCA); Greven *et al.* (2010) (LFPCA); Aston *et al.* (2010)]. Aston *et al.* (2010) projects the whole function onto a vector space, where the vector entries are the first few principal scores of the function. Through several linear mixed effect models which link PC scores and the covariates, they are able to assess the effect of covariates on the outcome function. Alternatively, MFPCA decomposes the intra-subject and inter-subject covariance operators in the two-way nested model, while inference is based on the scores separated by levels of variability. Using a similar approach, LFPCA models the longitudinal dynamics of functional observations measured at multiple visits. In this paper, we generalize these ideas to analyze functional observations collected under the most common nested and crossed designs, which expands the number and type of models for functional data. We propose structured functional principal component analysis (SFPCA) as a method to decompose the variability via PCA for any functional model with a particular linear structure. An attractive computational advantage of SFPCA is efficient handling of dense and high-frequency functional measurements. Throughout the paper, we only consider noise-free models because: 1) for low-dimensional data with white noise, the latent processes are estimable through smoothing; 2) for noise contaminated high-dimensional data, the problem is currently unresolved. To study the effect of this choice, we provide a simulation study in Section 4.2 for low-dimensional data with noise.

We organize the paper as follows: in Section 2, we provide a list of structured functional models that SFPCA is applied to and connect them to the symmetric sum method of moment (MoM) estimators described in Koch (1968); Section 3 discusses SFPCA and its implementation, with its extension to high-dimensional settings; Section 4 describes simulation studies for low-dimensional, high-dimensional and noisy settings; Section 5 applies SFPCA to the scientific problems described in the Introduction.

## 2 Structured Functional Models

Koch (1967) provided a comprehensive list of linear models for scalar data that emerge from various experimental designs. We contend that these models have natural extensions to functional data and that the models may be analyzed by decomposing the corresponding covariance operators. Table 1 includes the proposed designs that are grouped based on sampling schemes.

| | | |
|---|---|---|
| Nested | (N1) One-way | $Y_i(t) = \mu(t) + X_i(t)$ |
| | (N2) Two-way | $Y_{ij}(t) = \mu(t) + X_i(t) + U_{ij}(t)$ |
| | (N3) Three-way | $Y_{ijk}(t) = \mu(t) + X_i(t) + U_{ij}(t) + W_{ijk}(t)$ |
| | (NM) Multi-way | $Y_{i_1 i_2 \cdots i_r}(t) = \mu(t) + R^{(1)}_{i_1}(t) + R^{(2)}_{i_1 i_2}(t) + \cdots + R^{(r)}_{i_1 \cdots i_r}(t)$ |
| Crossed | (C2) Two-way | $Y_{ij}(t) = \mu(t) + X_i(t) + Z_j(t) + W_{ij}(t)$ |
| | (C2s) Two-way with subsampling | $Y_{ijk}(t) = \mu(t) + X_i(t) + Z_j(t) + W_{ij}(t) + U_{ijk}(t)$ |
| | (CM) Multi-way | $Y_{i_1 i_2 \cdots i_r u}(t) = \mu(t) + R_{\mathcal{T}_1}(t) + R_{\mathcal{T}_2}(t) + \cdots + R_{\mathcal{T}_d}(t)$ |

**Table 1:** *Structured functional models. For nested models, $i = 1, 2, \cdots, I; j = 1, 2, \cdots, J_i; k = 1, 2, \cdots, K_{ij}$; $i_1 = 1, 2, \cdots, I_1$, $i_2 = 1, 2, \cdots, I_{2i_1}, \ldots, i_r = 1, 2, \cdots, I_{r i_1 i_2 \ldots i_{r-1}}$. For crossed designs, $i = 1, 2, \cdots, I; j = 1, 2, \cdots, J; k = 1, 2, \cdots, n_{ij}$; (CM) contains combinations of any $s$ ($s = 1, 2, \cdots, r$) subset of the $r$ latent processes, as well as repeated measurements within each cell. $\mathcal{T}_1, \mathcal{T}_2, \cdots, \mathcal{T}_d \in \{i_{k_1} i_{k_2} \cdots i_{k_s} u : k_1, k_2, \cdots, k_s \in (1, 2, \cdots, r), u \in (\emptyset, 1, 2, \cdots, I_{i_1 i_2 \cdots i_r}), s \leq r\}$, $u$ is the index for repeated observation in cell $(i_{k_1}, i_{k_2}, \cdots i_{k_r})$.*

As shown in Table 1, the observed outcome, $Y(t)$, is expressed as a linear combination of latent processes $X(t)$, $U(t)$, $Z(t)$, and/or $W(t)$. We assume that these processes have mean zero and are square integrable, which guarantees their identifiability and mirrors standard statistical assumptions for scalar outcomes. Consequently, the total variability of a functional outcome is decomposed into process-specific variations. The covariance operators of latent processes define functional variance components of each model. Hence, these models capture a wide variety of correlation structures in modern functional data studies. We build up the intuition behind the functional nested and crossed designs, and connect them to the data examples that are discussed in the introduction.

## 2.1 Nested designs

A one-way nested model (N1) is the simplest variance component model for functional data. In (N1), the observed functional outcome $Y_i(t)$ is represented as a sum of a deterministic mean function, $\mu(t)$, and a level-specific stochastic process $X_i(t)$. $X_i(t)$ are assumed to be i.i.d, with mean zero and covariance operator $K_X(t, s) = \mathrm{E}\{X_i(t) X_i(s)\}$; $K_X$ may be thought of as the functional counterpart of scalar covariance. The variability of $Y_i(t)$ is completely determined by that of $X_i(t)$, that is, $K_Y = K_X$. In conventional functional data analysis (Ramsay & Silverman, 2005), $X_i(t)$ would be expressed via a set of spline or wavelet basis, or data-driven principal components [Ramsay & Silverman (2005); Di et al. (2009); Greven et al. (2010)]. Irrespective of the basis functions, $K_X$ is determined by the first two moments of the representation coefficients and a quadratic form of the basis functions.

The two-way functional nested design (N2) is the functional equivalent of a one-way analysis of variance (ANOVA) model. Originally motivated by the two-way sampling design of EEG data in SHHS (Di et al., 2009), the model expands (N1) with a subject-visit specific process $U_{ij}(t)$ that has covariance $K_U(t,s) = \text{Cov}\{U_{ij}(t), U_{ij}(s)\}$. Thus, the observed total variability of $Y_{ij}(t)$ is decomposed into subject-specific and subject-visit specific variability. These two parts are modeled through the corresponding functional covariance components $K_X$ and $K_U$ – covariance operators of $X_i(t)$ and $U_{ij}(t)$. To ensure identifiability, the random processes $X_i(t)$ and $U_{ij}(t)$ are assumed to have mean zero and be uncorrelated. This assumption also guarantees that $K_Y = K_X + K_U$.

Additional levels of nesting can be included in the model to accommodate corresponding hierarchies. For example, the three-way nested model (N3) provides an appropriate framework for modeling the activity intensity data described in Section 1. In addition to the subject-specific process $X_i(t)$ and the subject-visit specific process, $U_{ij}(t)$, the remaining part of the variation in $Y_{ijk}(t)$ is modeled through $W_{ijk}(t)$, which quantifies the hourly deviation from the average activity intensity level of day $j$ for subject $i$. The most general functional nested model (NM) admits arbitrarily many levels of nesting. If the activity intensity is followed for weeks or even months, a four-way or five-way model may be more appropriate, given the possibly repeated patterns of activity from week to week or month to month. As in the preceding models, mutual non-correlation is imposed for model identifiability. The total variability is decomposable into level-specific functional variance components as $K_Y = K_1 + K_2 + \cdots + K_r$, where $K_r(t,s) = \text{Cov}\{R^{(r)}_{i_1\cdots i_r}(t), R^{(r)}_{i_1\cdots i_r}(s)\}$. Here we used the notation from Table 1 for the multilevel model with an arbitrary number of levels (NM).

## 2.2 Crossed designs

Another group of designs admits crossing between levels. For example, the two-way crossed design (C2) is a functional analog of two-way ANOVA with an interaction term. It emphasizes a joint effect of two uncorrelated processes $X_i(t)$ and $Z_j(t)$, as well as their interaction $W_{ij}(t)$, on the outcome $Y_{ij}(t)$. The two-way crossed model with sub-sampling (C2s) applies to experimental designs where repeated measurements occur within each combination $(i, j)$ induced by the two first-level processes $X_i(t)$ and $Z_j(t)$. In addition to the first-level crossing $W_{ij}(t)$ as in (C2), $U_{ijk}(t)$ models the replications within each cell. For the phonetic analysis example, $X_i(t)$ and $Z_j(t)$ can model the main effects of syllable $i$ and speaker $j$, respectively, while $W_{ij}(t)$ models their interaction. Since multiple $F0$-contours may fall in category $(i, j)$, we use $U_{ijk}(t)$ to capture the remaining effect.

In general, we can consider an m-way crossed functional model (CM) with arbitrary number of crossings. In this model, $q$ ($q > 2$) uncorrelated latent processes have exchangeable first-level effects on $Y(t)$. Any subset of $s$ ($s \leq q$) processes out of $q$ may have interactions, resulting in $d$ functional additive terms in the model. For notational convenience, we express this model using $d$ sub-index sets, $\{\mathcal{T}_1, \mathcal{T}_2, \ldots, \mathcal{T}_d\}$ that define the model structure. For example, (C2s) with four terms can be written as $\{\mathcal{T}_1, \mathcal{T}_2, \mathcal{T}_3, \mathcal{T}_4\} = \{i, j, ij, ijk\}$ and $R_{\mathcal{T}_1}(t) = X_i(t)$, $R_{\mathcal{T}_2}(t) = Z_j(t)$, $R_{\mathcal{T}_3}(t) = W_{ij}(t)$, and $R_{\mathcal{T}_4}(t) = U_{ijk}(t)$. The assumptions on correlation structure stay the same; all latent processes, including interaction terms, are uncorrelated. We now show how to efficiently estimate these models.

## 3 Structured Functional PCA

We develop structured functional PCA (SFPCA) to efficiently reduce dimensionality and extract signals for the class of functional models introduced in Section 2. This approach models latent processes parsimoniously via principal components by Karhunen-Loéve expansion [Karhunen (1947); Loève (1978)]. SFPCA starts with estimating the covariance operators of latent processes. Following Koch (1968), we employ the MoM approach based on symmetric sums. By extending his approach to functional settings, we construct unbiased estimators of covariance matrices on a grid of $p$ points $\mathcal{T} = \{t_1, t_2, \ldots, t_p\}$. After estimating the covariance operators, we conduct spectral decomposition to obtain principal components and principal scores that serve as coordinates in the space spanned by principal components.

We use two-way crossed design (C2) as the main example. Details for other models in Table 1 can be found in the Appendix. Let $\mu(t)$ be the fixed population mean, $X_i(t)$, $Z_j(t)$ and $W_{ij}(t)$ be mutually uncorrelated mean-zero random processes as described in Section 2. Their covariance operators are $K_X$, $K_Z$ and $K_W$, respectively, where $K_X(t,s) = \mathrm{E}\{X_i(t)X_i(s)\}$, $K_Z(t,s) = \mathrm{E}\{Z_j(t)Z_j(s)\}$ and $K_W(t,s) = \mathrm{E}\{W_{ij}(t)W_{ij}(s)\}$. Using the Karhunen-Loéve expansion for $X_i(t), Z_j(t)$, and $W_{ij}(t)$, model (C2) becomes

$$Y_{ij}(t) = \mu(t) + \sum_{k=1}^{\infty} \phi_k^X(t)\xi_{ik}^X + \sum_{l=1}^{\infty} \phi_l^Z(t)\xi_{jl}^Z + \sum_{m=1}^{\infty} \phi_m^W(t)\xi_{ijm}^W, \quad (1)$$

where $\phi_k^X(t)$, $\phi_l^Z(t)$ and $\phi_m^W(t)$ are the eigenfunctions of the covariance operators $K_X$, $K_Z$ and $K_W$, respectively. The scores $\xi_{ik}^X = \int X_i(s)\phi_k^X(s)ds$, $\xi_{il}^Z = \int Z_j(s)\phi_l^Z(s)ds$, and $\xi_{ijm}^W = \int W_{ij}(s)\phi_m^W(s)ds$ are mutually independent random variables with mean 0 and variance $\lambda_k^X$, $\lambda_l^Z$, and $\lambda_m^W$, respectively, where $\lambda_k^X \geq \lambda_{k+1}^X$, $\lambda_l^Z \geq \lambda_{l+1}^Z$, and $\lambda_m^W \geq \lambda_{m+1}^W$ for every $k$, $l$, and $m$. Normality of scores is not necessary for the results in this paper, but may be a convenient mild assumption.

## 3.1 Principal scores estimation

Consider the case when most variability of each latent process is captured by the first $N_1$, $N_2$ and $N_3$ principal components of $X_i(t)$, $Z_j(t)$ and $W_{ij}(t)$, respectively, model (1) can then be approximated as $Y_{ij}(t) = \mu(t) + \sum_{k=1}^{N_1} \phi_k^X(t)\xi_{ik}^X + \sum_{l=1}^{N_2} \phi_l^Z(t)\xi_{jl}^Z + \sum_{m=1}^{N_3} \phi_m^W(t)\xi_{ijm}^W$. Let $\mathbf{Y} =$

$(\mathbf{Y}_{11},\ldots,\mathbf{Y}_{1J_1},\cdots,\mathbf{Y}_{I1},\ldots,\mathbf{Y}_{IJ_I})$ be a $p \times n$ matrix with $\mathbf{Y}_{ij} := \{Y_{ij}(t_1), Y_{ij}(t_2), \cdots, Y_{ij}(t_p)\}^T$ and $n = \sum_{i=1}^{I} J_i$. For notational simplicity we assume a balanced design where $J_i = J$, though the assumption is not necessary. Let $\boldsymbol{\xi}_i^X = (\xi_{i1}^X, \cdots, \xi_{iN_1}^X)^T$ and $\boldsymbol{\Phi}_X = [\boldsymbol{\phi}_1^X(\mathcal{T}), \boldsymbol{\phi}_2^X(\mathcal{T}), \cdots, \boldsymbol{\phi}_{N_1}^X(\mathcal{T})]$ be the first $N_1$ principal components observed at time grid $\mathcal{T}$. Similar definition applies to $(\boldsymbol{\xi}_j^Z, \boldsymbol{\Phi}_Z)$ and $(\boldsymbol{\xi}_{ij}^W, \boldsymbol{\Phi}_W)$. Hence the truncated model is further expressed into matrix form as $\mathbf{Y}_{ij} = \boldsymbol{\mu} \otimes \mathbf{1}_p^T + \boldsymbol{\Phi}_X \boldsymbol{\xi}_i^X + \boldsymbol{\Phi}_Z \boldsymbol{\xi}_j^Z + \boldsymbol{\Phi}_W \boldsymbol{\xi}_{ij}^W$.

We will show in the next section how to obtain $\widehat{K}_X$, $\widehat{K}_Z$ and $\widehat{K}_W$. Assume now that such estimators are available and let $\widehat{\boldsymbol{\Phi}}_X^{N_1}$, $\widehat{\boldsymbol{\Phi}}_Z^{N_2}$ and $\widehat{\boldsymbol{\Phi}}_W^{N_3}$ be their first $N_1$, $N_2$ and $N_3$ eigenvectors, respectively. Denote the vectors of first $N_1$, $N_2$ and $N_3$ eigenvalues for the covariance matrices as $\widehat{\boldsymbol{\Lambda}}_X^{N_1}$, $\widehat{\boldsymbol{\Lambda}}_Z^{N_2}$ and $\widehat{\boldsymbol{\Lambda}}_W^{N_3}$. We can estimate the truncated set of principal scores to be the best linear unbiased predictor (BLUP) of the mixed effect model $\mathbf{Y}_{ij} = \widehat{\boldsymbol{\mu}} \otimes \mathbf{1}_p^T + \widehat{\boldsymbol{\Phi}}_X^{N_1} \boldsymbol{\xi}_i^X + \widehat{\boldsymbol{\Phi}}_Z^{N_2} \boldsymbol{\xi}_j^Z + \widehat{\boldsymbol{\Phi}}_W^{N_3} \boldsymbol{\xi}_{ij}^W$, where $\boldsymbol{\xi}_i^X \sim N(\mathbf{0}, \widehat{\boldsymbol{\Lambda}}_X^{N_1})$, $\boldsymbol{\xi}_j^Z \sim N(\mathbf{0}, \widehat{\boldsymbol{\Lambda}}_Z^{N_2})$ and $\boldsymbol{\xi}_{ij}^W \sim N(\mathbf{0}, \widehat{\boldsymbol{\Lambda}}_W^{N_3})$; $\widehat{\boldsymbol{\mu}}$ may be obtained by sample average. Specifically, for the two-way crossed model (C2) and three-way nested model (N3), their estimated BLUPs are provided in Appendix A.

## 3.2 MoM covariance operator estimation

By extending the idea of symmetric sum MoM estimators in Koch (1968), we show that our estimated covariance matrices will be of the form $\widehat{K}_X = \mathbf{Y}\mathbf{G}_X\mathbf{Y}^T$, $\widehat{K}_Z = \mathbf{Y}\mathbf{H}_Z\mathbf{Y}^T$ and $\widehat{K}_W = \mathbf{Y}\mathbf{G}_W\mathbf{Y}^T$, where $\mathbf{G}_X$, $\mathbf{G}_Z$ and $\mathbf{G}_W$ are specific matrices of dimension $n \times n$. In fact, for all the structured functional models, MoM estimators of covariance operator is representable in the "sandwich" form, $\mathbf{Y}\mathbf{G}\mathbf{Y}^T$. We illustrate the detailed calculation for the covariance operators for the two-way crossed design (C2) and three-way nested design (N3). Results for other design schemes are provided in Appendix B. We start by subtracting an estimator of the mean $\mu(t)$ and hence assume $\mu(t) = 0$ from now on.

### 3.2.1 Two-way crossed design (C2)

For model (C2), we have

$$E\{Y_{ij}(t) - Y_{kl}(t)\}\{(Y_{ij}(s) - Y_{kl}(s))\}^T = \begin{cases} 2\{K_W(t,s) + K_Z(t,s)\}, & \text{if } i = k, j \neq l \\ 2\{K_W(t,s) + K_X(t,s)\}, & \text{if } i \neq k, j = l \\ 2\{K_W(t,s) + K_Z(t,s) + K_X(t,s)\}, & \text{if } i \neq k, j \neq l \end{cases}$$

Let $n_{ij} = 1$ if $Y_{ij}$ is observed and 0 otherwise; $n_{i0} = \sum_i n_{ij}$, $n_{0j} = \sum_j n_{ij}$, $n = \sum_{i,j} n_{ij}$, $k_1 = \sum_i n_{i0}^2$ and $k_2 = \sum_j n_{0j}^2$. Define $\mathbf{D}_{n \times n} = \text{diag}\{\mathbf{N}_1, \mathbf{N}_2, \cdots, \mathbf{N}_I\}$ with $\mathbf{N}_i = n_{i0}\mathbf{I}_{n_{i0}}$, $\mathbf{E}_{I \times n} = \text{diag}\{\mathbf{1}_{n_{10}}^T, \cdots, \mathbf{1}_{n_{I0}}^T\}$ and $\mathbf{1}_n = (1, 1, \cdots, 1)^T$. $\mathbf{P}_{n \times n} = \text{diag}\{\mathbf{P}_1, \cdots, \mathbf{P}_I\}$ with $\mathbf{P}_i = \text{diag}\{n_{01}, \cdots, n_{0n_{i0}}\}$ of dimension $n_{i0} \times n_{i0}$, $\mathbf{F}_{J \times n} = (\mathbf{f}_1, \cdots, \mathbf{f}_J)^T$ is the second-level analogy to $\mathbf{E}$, where $\mathbf{f}_j$ is a vector with value 1 on observations with second-level process $Z_j(t)$ and 0 otherwise. If $H_Z = 2(K_W + K_Z)$, $H_X = 2(K_W + K_X)$, and $H_{XZ} = 2(K_W + K_Z + K_X)$ then using the results above we obtain the following *explicit* MoM estimators

$$\widehat{H}_Z = \frac{1}{k_1 - n} \sum_{i=1} \sum_{j \neq l} (\mathbf{Y}_{ij} - \mathbf{Y}_{il})(\mathbf{Y}_{ij} - \mathbf{Y}_{il})^T = \frac{2}{k_1 - n} \mathbf{Y}(\mathbf{D} - \mathbf{E}^T\mathbf{E})\mathbf{Y}^T$$

$$\widehat{H}_X = \frac{1}{k_2 - n} \sum_{i \neq k} \sum_j (\mathbf{Y}_{ij} - \mathbf{Y}_{kj})(\mathbf{Y}_{ij} - \mathbf{Y}_{kj})^T = \frac{2}{k_2 - n} \mathbf{Y}(\mathbf{P} - \mathbf{F}^T\mathbf{F})\mathbf{Y}^T$$

$$\widehat{H}_{XZ} = \frac{1}{n^2 - k_1 - k_2 + n} \sum_{i \neq k} \sum_{j \neq l} (\mathbf{Y}_{ij} - \mathbf{Y}_{kl})(\mathbf{Y}_{ij} - \mathbf{Y}_{kl})^T$$

$$= \frac{2}{n^2 - k_1 - k_2 + n} \mathbf{Y}(n\mathbf{I} - \mathbf{1}\mathbf{1}^T - \mathbf{D} + \mathbf{E}^T\mathbf{E} - \mathbf{P} + \mathbf{F}^T\mathbf{F})\mathbf{Y}^T$$

Thus, the covariance operators can be estimated as $\widehat{K}_Z = (\widehat{H}_{XZ} - \widehat{H}_X)/2 = \mathbf{Y}\mathbf{G}_Z\mathbf{Y}^T$, $\widehat{K}_X = (\widehat{H}_{XZ} - \widehat{H}_Z)/2 = \mathbf{Y}\mathbf{G}_X\mathbf{Y}^T$ and $\widehat{K}_W = (\widehat{H}_X + \widehat{H}_Z - \widehat{H}_{XZ})/2 = \mathbf{Y}\mathbf{G}_W\mathbf{Y}$, where $\mathbf{G}_Z = \frac{1}{n^2 - k_1 - k_2 + n}\{n\mathbf{I} - \mathbf{1}\mathbf{1}^T - \mathbf{D} + \mathbf{E}^T\mathbf{E} - \frac{n^2 - k_1}{k_2 - n}(\mathbf{P} - \mathbf{F}^T\mathbf{F})\}$, $\mathbf{G}_X = \frac{1}{n^2 - k_1 - k_2 + n}\{n\mathbf{I} - \mathbf{1}\mathbf{1}^T - \mathbf{P} + \mathbf{F}^T\mathbf{F} - \frac{n^2 - k_2}{k_1 - n}(\mathbf{D} - \mathbf{E}^T\mathbf{E})\}$ and $\mathbf{G}_W = \frac{1}{n^2 - k_1 - k_2 + n}\{\frac{n^2 - k_1}{k_2 - n}(\mathbf{P} + \mathbf{F}^T\mathbf{F}) + \frac{n^2 - k_2}{k_1 - n}(\mathbf{D} - \mathbf{E}^T\mathbf{E}) - n\mathbf{I} + \mathbf{1}\mathbf{1}^T\}$.

### 3.2.2 Three-way nested model

Consider now model (N3), where $Y_{ijk}(t) = X_i(t) + U_{ij}(t) + W_{ijk}(t)$, $i = 1, 2, \cdots, I$; $j = 1, 2, \cdots, J_i$; $k = 1, 2, \cdots, n_{ij}$, and $X$, $U$ and $W$ to be the three latent processes nested within each other. Similar to the approach for (C2), we have

$$E\{Y_{ijk}(t) - Y_{luv}(t)\}\{Y_{ijk}(s) - Y_{luv}(s)\}^T = \begin{cases} 2K_W(t,s), & \text{if } i = l, j = u, k \neq v \\ 2\{K_W(t,s) + K_U(t,s)\}, & \text{if } i = l, j \neq u \\ 2\{K_X(t,s) + K_W(t,s) + K_U(t,s)\}, & \text{if } i \neq l \end{cases}$$

Again let $\mathbf{Y}_{ijk} := \{Y_{ijk}(t_1), \cdots, Y_{ijk}(t_p)\}^T$, $n := \sum_{i,j}^I n_{ij}$, $n_{i\cdot} = \sum_j^{J_i} n_{ij}$, $k_1 = \sum_{i=1}^I \sum_{j=1}^{J_i} n_{ij}^2$, $k_2 = \sum_i n_{i\cdot}^2$. $\mathbf{D}_1 = \text{diag}\{N_{11}, \cdots, N_{I m_I}\}$ and $\mathbf{D}_2 = \text{diag}\{N_1, \cdots, N_I\}$, where $N_{ij} = n_{ij} I_{n_{ij}}$ and $N_i = n_{i\cdot} I_{n_{i\cdot}}$; $\mathbf{E}_1 = \text{diag}\{\mathbf{1}_{n_{11}}^T, \cdots, \mathbf{1}_{n_{I m_I}}^T\}$, $\mathbf{E}_2 = \text{diag}\{\mathbf{1}_{n_{1\cdot}}^T, \cdots, \mathbf{1}_{n_{I\cdot}}^T\}$. If $H_W = 2K_W$, $H_U = 2(K_W + K_U)$, and $H_x = 2(K_W + K_Z + K_X)$ then using the results above we obtain the following MoM estimators

$$\widehat{H}_W = \frac{1}{k_1 - n} \sum_{i,j} \sum_{k,v} (Y_{ijk} - Y_{ijv})(Y_{ijk} - Y_{ijv})^T = \frac{2}{k_1 - n} \mathbf{Y}(\mathbf{D}_1 - \mathbf{E}_1^T \mathbf{E}_1) \mathbf{Y}^T$$

$$\widehat{H}_U = \frac{1}{k_2 - k_1} \sum_i \sum_{j \neq u} \sum_{k,v} (Y_{ijk} - Y_{iuv})(Y_{ijk} - Y_{iuv})^T = \frac{2}{k_2 - k_1} \mathbf{Y}(\mathbf{D}_2 - \mathbf{E}_2^T \mathbf{E}_2 - \mathbf{D}_1 + \mathbf{E}_1^T \mathbf{E}_1) \mathbf{Y}^T$$

$$\widehat{H}_X = \frac{1}{n^2 - k_2} \sum_{i \neq l} \sum_{j,u,k,v} (Y_{ijk} - Y_{luv})(Y_{ijk} - Y_{luv})^T = \frac{2}{n^2 - k_2} \mathbf{Y}(n\mathbf{I} - \mathbf{1}\mathbf{1}^T - \mathbf{D}_2 + \mathbf{E}_2^T \mathbf{E}_2) \mathbf{Y}^T$$

Then $\widehat{K}_W = \widehat{H}_W/2$, $\widehat{K}_U = (\widehat{H}_U - \widehat{H}_W)/2$ and $\widehat{K}_X = (\widehat{H}_X - \widehat{H}_U)/2$ all have the form $\mathbf{Y}\mathbf{G}\mathbf{Y}^T$.

In general, multi-way nested and crossed designs can be estimated through a similar work flow (see Appendix B for details).

## 3.3 Structured high-dimensional data

Given the current research emphasis on high-dimensional data, linear models are still considered to be difficult to fit for high-dimensional data. Here we show that the entire class of models described in Table 1 can be fit using fast approaches. Note that the estimation procedures in the previous sections assume that the MoM estimators of the relevant covariance operators can be constructed and decomposed. When the dimension of observations, $p$, is moderate, the methods described in Section 3 are straightforward. However, if the observations are high or ultra-high dimensional, the approach is no longer feasible. Because calculating and storing a $p$-dimensional covariance operator $\widehat{K}_{p \times p}$ is computationally expen-sive, while its spectral decomposition becomes prohibitive. Thus, we propose an alternative approach based on a rank-preserving argument. This algorithm allows efficient calculation of the eigenvectors and eigenvalues without requiring either storing or diagonalizing the estimated high-dimensional covariance operators. The algorithm is outlined below.

Throughout this section, we assume that $p \gg n$. Hence, the induced covariance matrix is at most of rank $n$. Zipunnikov *et al.* (2011) proposed an approach that avoids calculating the covariance operators in the original high-dimensional space. Consider (C2) as an example: the idea is to map the model onto a lower-dimensional space and obtain $\widetilde{\mathbf{Y}}_{ij} := \mathbf{C}\mathbf{Y}_{ij} = \mathbf{C}\boldsymbol{\Phi}^X\boldsymbol{\xi}_i^X + \mathbf{C}\boldsymbol{\Phi}^Z\boldsymbol{\xi}_j^Z + \mathbf{C}\boldsymbol{\Phi}^W\boldsymbol{\xi}_{ij}^W$. The matrix $\mathbf{C}$ can be anything as long as the left (outer) dimension is small and the right (inner) dimension is equal to the dimension of the problem. One possible choice of $\mathbf{C}$ would be to start with the whole data matrix, $\mathbf{Y}$, which can be obtained by column binding individual data vectors, $\mathbf{Y}_{ij}$. If $\mathbf{Y} = \mathbf{V}\mathbf{S}^{1/2}\mathbf{U}^T$ is the singular value decomposition of $\mathbf{Y}$, then we can simply choose $\mathbf{C} = \mathbf{V}^T$. The model becomes $\mathbf{S}^{1/2}\mathbf{U}^T = \mathbf{V}^T\boldsymbol{\Phi}^X\boldsymbol{\xi}_i^X + \mathbf{V}^T\boldsymbol{\Phi}^Z\boldsymbol{\xi}_j^Z + \mathbf{V}^T\boldsymbol{\Phi}^W\boldsymbol{\xi}_{ij}^W := \mathbf{A}^X\boldsymbol{\xi}_i^X + \mathbf{A}^Z\boldsymbol{\xi}_j^Z + \mathbf{A}^W\boldsymbol{\xi}_{ij}^W$. Theorem 1 in Zipunnikov *et al.* (2011) shows that this transformation preserves full information for the linear PCA model. The eigenfunctions for the original model can be recovered by left multiplying $\mathbf{V}$ to those obtained in the new model and the eigenvalues remain unchanged. This is straightforward to implement, as the number of operations involved in calculating the SVD of $\mathbf{Y}$ is linear in the dimension of the data, $p$. After obtaining the SVD of $\mathbf{Y}$, each column $\mathbf{Y}_{ij}$ can be represented as $\mathbf{Y}_{ij} = \mathbf{V}\mathbf{S}^{1/2}\mathbf{U}_{ij}$, where $\mathbf{U}_{ij}$ is a corresponding column of matrix $\mathbf{U}^T$. Therefore, the vectors $\mathbf{Y}_{ij}$ differ only via the factors $\mathbf{U}_{ij}$ of length $n$, which is much lower-dimensional. Comparing this SVD representation of $\mathbf{Y}_{ij}$ with the original model (C2), it follows that the structured separation of the variability modeled by high-dimensional latent processes $X_i$, $Z_j$, and $W_{ij}$ is identically in the structured separation of the low-dimensional vectors $\mathbf{U}_{ij}$. This is the key observation which motivates our approach. This model has an "intrinsic" dimensionality that is induced by the sample size $n$. This low-dimensional model is estimable using SFPCA in Section 3 and requires only $O(n^3)$ calculations.

After obtaining $\mathbf{A}_{N_1}^X$, $\mathbf{A}_{N_2}^Z$ and $\mathbf{A}_{N_3}^W$ as the estimates for $\mathbf{A}^X$, $\mathbf{A}^Z$ and $\mathbf{A}^W$, we can obtain $\tilde{\boldsymbol{\xi}}_i^X$, $\tilde{\boldsymbol{\xi}}_j^Z$ and $\tilde{\boldsymbol{\xi}}_{ij}^W$ as induced BLUP in the lower-dimensional model, with the matrices $\mathbf{A}$'s replaced by their corresponding estimates. Furthermore, $\boldsymbol{\Phi}_{N_1}^X$, $\boldsymbol{\Phi}_{N_2}^Z$ and $\boldsymbol{\Phi}_{N_3}^W$ in the original space may be recovered by left multiplying $\mathbf{V}$ onto $\mathbf{A}_{N_1}^X$, $\mathbf{A}_{N_2}^Z$ and $\mathbf{A}_{N_3}^W$. We provide the formula for final estimates and their detailed derivation of two-way crossed model (C2) and three-way nested model (N3) in Appendix A. Up until the last step, all the calculations can be conducted in $O(n^3)$ complexity. Therefore, fitting the model in a reduced-dimensional space guarantees the high-dimensional principal components in a p-linear time. This means that complex statistical models for ultra-high dimensional data sets can be fitted quickly.

## 4 Simulations

To better understand how SFPCA performs in practice, we consider both low- and high-dimensional simulation scenarios. We conduct simulation studies for the two-way crossed design (C2) and three-way nested model (N3), and evaluate the estimation under various signal-to-noise ratios.

### 4.1 Models without white noise

For the two-way crossed design (C2), we generate data from the following model

$$\begin{cases} Y_{ij}(t) = \sum_{k=1}^{N_X} \phi_k^X(t)\xi_{ik} + \sum_{l=1}^{N_Z} \phi_l^Z(t)\zeta_{jl} + \sum_{h=1}^{N_W} \phi_h^W(t)\eta_{ijh}, t \in \mathcal{T} \\ \xi_{ik} \stackrel{i.i.d}{\sim} N(0, \lambda_k^X), \zeta_{jl} \stackrel{i.i.d}{\sim} N(0, \lambda_l^Z) \text{ and } \eta_{ijh} \stackrel{i.i.d}{\sim} N(0, \lambda_h^W) \end{cases} \quad (2)$$

where $\xi_{ik}$'s, $\zeta_{jl}$'s and $\eta_{ijh}$'s are mutually uncorrelated. We choose $N_X = N_Z = N_W = 4$ and let the true eigenvalues be $\lambda_k^X = \lambda_k^Z = \lambda_k^W = 0.5^{k-1}, k = 1, 2, 3, 4$. True eigenfunctions are

$$\phi_1^X(t) = \sin(2\pi t) \quad \phi_2^X(t) = \cos(2\pi t) \quad \phi_3^X(t) = \sin(4\pi t) \quad \phi_4^X(t) = \cos(4\pi t)$$

$$\phi_1^Z(t) = \sin(6\pi t) \quad \phi_2^Z(t) = \cos(6\pi t) \quad \phi_3^Z(t) = \sin(8\pi t) \quad \phi_4^Z(t) = 1/\sqrt{2}$$

$$\phi_1^W(t) = \sqrt{3}(2t-1) \quad \phi_2^W(t) = \sqrt{5}(6t^2 - 6t + 1) \quad \phi_3^W(t) = \sqrt{5}(20t^3 - 30t^2 + 12t - 1) \quad \phi_4^W(t) = 1$$

These functions are measured on the grid $\mathcal{T} = 1/p, 2/p, \cdots, 1$. For the low-dimensional case, we choose $p = 100$. Let $I = 200$ and $J = 20$ be the number of categories for first-level processes $X(t)$ and $Z(t)$, respectively. Figure 3 shows an example of the simulated curves.

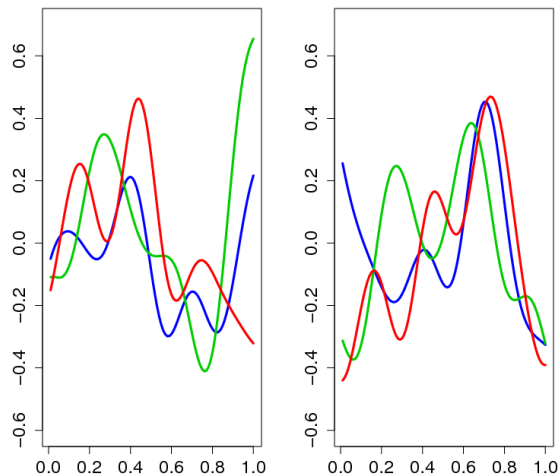

**Figure 3:** *An example of the simulated functional data following model (C2) as listed in equation (2), with parameters specified in the text. The three curves within each panel share the same $X(t)$, but vary on $Z(t)$ and the interaction term $W(t)$; curves with the same color across two panels share the same $Z(t)$, but may vary on others. Similarities within each panel as well as on second-level effect is observable.*

We repeat simulation for 100 times and display the estimated principal components for the three latent processes in Figure 4. The procedure recovers $X(t)$ and $W(t)$ very well. Given the small number of samples that is observed for $Z(t)$, estimation for $K_Z$ and therefore its eigenfunctions are more noisy and unstable. This can be resolved by increasing $J$. In fact, we examine another two scenarios where $I = 20$, $J = 200$ (upper panel of Figure 9 in Appendix C) and $I = J = 200$ (bottom panel) to evaluate the effect of sample size on recovering each variance component. The reliability in estimating the eigenfunctions increases when there are more observations per latent process.

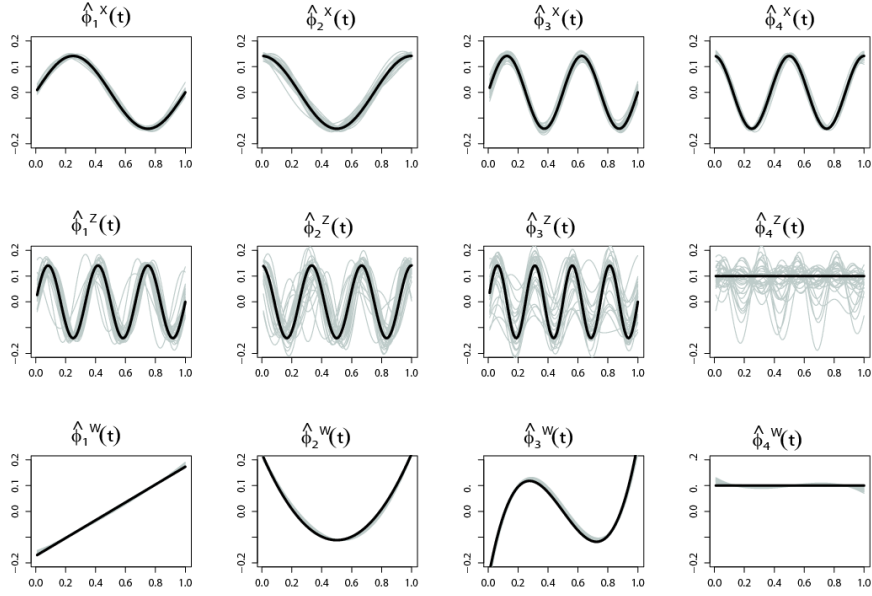

**Figure 4:** *The estimated principal components when $I = 200$ and $J = 20$ in 100 simulations are shown in gray bands. Black curves are the true eigenfunctions.*

For the three-way nested model (N3), we study the high-dimensional scenario using a similar parameterization with example. Consider

$$\begin{cases} Y_{ijk}(t) = \sum_{l=1}^{N_1} \phi_l^X(t)\xi_{il} + \sum_{m=1}^{N_2} \phi_m^U(t)\zeta_{ijm} + \sum_{h=1}^{N_3} \phi_h^W(t)\eta_{ijkh}, t \in \mathcal{T} \\ \xi_{il} \overset{i.i.d}{\sim} N(0, \lambda_l^{(1)}), \zeta_{ijm} \overset{i.i.d}{\sim} N(0, \lambda_m^{(2)}) \text{ and } \eta_{ijkh} \overset{i.i.d}{\sim} N(0, \lambda_h^{(3)}) \end{cases} \quad (3)$$

where $U_{ij}(t)$'s ($j = 1, 2, \cdots, J$) are nested within $i$, and $W_{ijk}(t)$'s ($k = 1, 2, \cdots, K$) are nested within $ij$. $\xi_{il}$'s, $\zeta_{ijm}$'s and $\eta_{ijkh}$'s are assumed to be uncorrelated.

Let $N_1 = N_2 = N_3 = 4$, $\lambda_k^{(1)} = \lambda_k^{(2)} = \lambda_k^{(3)} = 0.5^{k-1}, k = 1, 2, 3, 4.$, and the true eigenfunctions

$\phi_1^X(t) = \sin(2\pi t) \quad \phi_2^X(t) = \cos(2\pi t) \quad \phi_3^X(t) = \sin(4\pi t) \quad \phi_4^X(t) = \cos(4\pi t)$

$\phi_1^U(t) = \sin(6\pi t) \quad \phi_2^U(t) = \cos(6\pi t) \quad \phi_3^U(t) = \sin(8\pi t) \quad \phi_4^U(t) = 1/\sqrt{2}$

$\phi_1^W(t) = 1 \quad \phi_2^W(t) = \sqrt{3}(2t - 1) \quad \phi_3^W(t) = \sqrt{5}(6t^2 - 6t + 1) \quad \phi_4^W(t) = \sqrt{5}(20t^3 - 30t^2 + 12t - 1)$

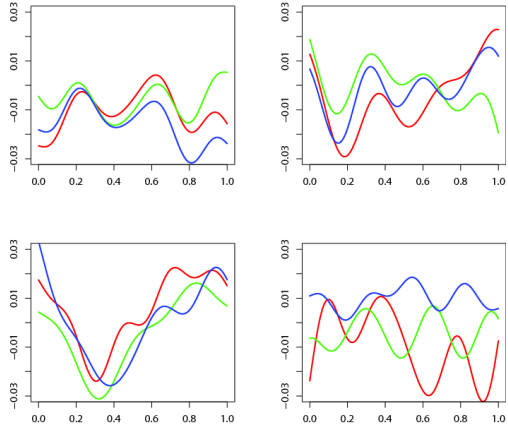

**Figure 5:** *Simulated functions following the three-way nested model as listed in equation (3). The three curves in the upper left panel are random samples that share the same first- and second-level processes. Their correlations across curves are induced by both $X(t)$ and $U(t)$. On the upper-right panel are samples with the same first-level process $X(t)$ only, but vary on the second- and third-level process $U(t)$ and $W(t)$. The correlation among these five curves is much weaker than the previous plot. The bottom-left plot are those share the same first- and third-level processes $X(t)$ and $W(t)$, but vary on the second-level process $Z(t)$. Finally, the bottom-right plot shows three completely uncorrelated curves from the model.*

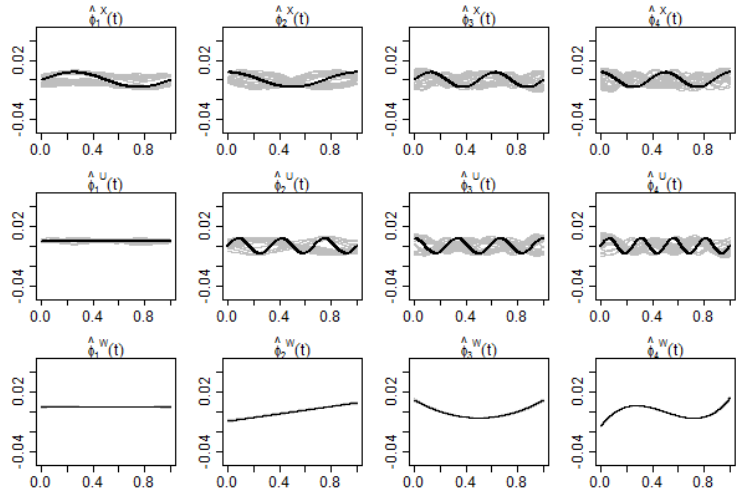

**Figure 6:** *The estimated principal components under high-dimensional setting over 100 simulations are shown in gray (we randomly plot 50 out of 100 estimates). The true eigenfunctions are displayed in black curves. The first- and second-level hierarchies are captured by two sets of trigonometric basis. The third-level processes is polynomial. We deliberately use a thinner line width for the black curves on the third row, so that the estimated PCs in gray are still visible.*

With $p = 35,000$, $I = 50$, $J = 5$ and $K = 5$, we display the simulated data under various nested structures in Figure 5 and the estimated principal components in Figure 6. Note that the estimated eigenfunctions are in fact the linear combinations of eigenfunctions from other levels. For instance, the estimates for constant functions $\phi_1^U(t)$ and $\phi_1^W(t)$ demonstrate trigonometric shape. The deviations, although small on the scale of the functions, can diminish as sample size increases.

## 4.2 Models with white noise

So far have assumed that data are measured without noise, or that noise has a smooth covariance structure that can be absorbed into one of the latent processes. However, SFPCA can easily be extended to data contaminated by white noise $\epsilon \sim N(0, \sigma^2)$. The only difference is that the new symmetric sum operators $\widetilde{H}$ have larger diagonal values. i.e., $\widetilde{H}(t,s) = H(t,s) + \sigma^2 \delta_{ts}$, where $\delta_{ts}$ is the Dirichlet function. We estimate $H(t,s)$ by smoothing the off-diagonal surface of $\widetilde{H}$ as, for example, in Staniswalis & Lee (1998). For high-dimensional case, although the white noise remains $i.i.d$ with the same variance after left multiplying with an orthonormal matrix as described in Section 3.3, we encounter multiple difficulties. Therefore, we do not consider high-dimensional case with noise for the time being.

Specifically, for model (N3), we have $\widetilde{K}_W(t,s) = K_W(t,s) + \sigma^2 \delta_{ts}$, $\widetilde{K}_U(t,s) = K_U(t,s)$, $\widetilde{K}_X(t,s) = K_X(t,s)$. Thus, we only need to smooth the off-diagonal elements of $\widetilde{K}_W$ to estimate $\sigma^2$. We show a simulation example of SFPCA for the N3 model under the same setting as in the previous section. The mean square errors (MSEs) of the estimated principal components and white noise variance are in Table 2.

To better summarize the signal-to-noise ratio, we define reliability ratio (Shou *et al.*, 2012) to be the ratio of variances across all observed time points between processes $X(t)$, $U(t)$ and $W(t)$, and the observed outcome $RR = \frac{\text{trace}\{K_X + K_U + K_W\}}{\text{trace}\{K_X + K_U + K_W + \sigma^2\}}$. Similar summary statistics can be defined for other models listed in Table 1.

|  | $\sigma_\eta^2 = 0$ | $\sigma_\eta^2 = 0.25$ | $\sigma_\eta^2 = 0.5$ | $\sigma_\eta^2 = 1$ |
|---|---|---|---|---|
| $\hat{\sigma}^2$ | 2.38e-09 | 8.63e-04 | 4.58e-06 | 1.97e-05 |
| $\phi_1^X$ | 6.55e-04 | 8.63e-04 | 1.24e-03 | 1.60e-03 |
| $\phi_2^X$ | 1.65e-03 | 2.07e-03 | 2.64e-03 | 3.30e-03 |
| $\phi_3^X$ | 2.10e-03 | 2.99e-03 | 3.68e-03 | 6.04e-03 |
| $\phi_4^X$ | 2.63e-03 | 6.33e-03 | 8.12e-03 | 1.15e-02 |
| $\phi_1^U$ | 1.84e-04 | 5.37e-04 | 8.27e-04 | 1.50e-03 |
| $\phi_2^U$ | 2.93e-04 | 9.46e-04 | 1.58e-03 | 3.14e-04 |
| $\phi_3^U$ | 2.91e-04 | 1.55e-03 | 3.18e-03 | 7.59e-03 |
| $\phi_4^U$ | 2.25e-04 | 3.01e-03 | 7.76e-03 | 1.54e-02 |
| $\phi_1^W$ | 2.82e-05 | 3.65e-04 | 7.92e-04 | 2.07e-03 |
| $\phi_2^W$ | 7.46e-05 | 8.51e-04 | 2.16e-03 | 6.80e-03 |
| $\phi_3^W$ | 1.45e-04 | 2.10e-03 | 6.65e-03 | 1.55e-02 |
| $\phi_4^W$ | 1.81e-04 | 6.60e-03 | 1.59e-02 | 1.76e-02 |

**Table 2:** *Average MSE of the first 4 principal components under different signal-to-noise ratio.*

# 5 Data Applications

SFPCA can be applied to various types of structured data including the three examples discussed in the introduction. The SHHS data was analyzed in details in Di *et al.* (2009) with MFPCA, which is a special case of the methodology considered in this paper. Here we provide results for the phonetic study and the accelerometer data.

## 5.1 Phonetic study

The phonetic study of Luobuzhai Qiang dialect, as described in the Introduction, consists of F0-contours from 8 subjects speaking 19 words. Each word contains 3 to 12 different vowels. There are 5 different vowels: 'ə', 'a', 'e', 'i' and 'u'. For each vowel within a particular word, pitch is measured at 11 equidistant time points, with time standardized by the length of the vowel. This dataset is not high-dimensional, but is treated as a function because it was assessed to be smooth (Aston *et al.*, 2010). To assess the effect of covariates with relatively simple specification, Aston *et al.* (2010) assumes the same principal components for all latent processes. Covariates only influence the pitch curves through principal scores – weights of each principal components. Here we relax these assumptions and attempt to fully evaluate the variability and effect of each latent process that is indicated by the data structure. Because each word was repeated under three different contexts by each speaker, we take a two-way crossed with sub-sampling (C2s) model as in Table 1; we absorb $W_{ij}(t)$ into $U_{ijk}(t)$ because the sample size is not large enough to estimate $W_{ij}(t)$ accurately. In fact, our model can be written as $Y_{ijk}(t) = \mu(t) + X_i(t) + Z_j(t) + U_{ijk}(t)$, with the first-level random effect $X_i(t), i = 1, 2, \cdots, 8$ accounting for variation among speakers. We choose the second-level random effect $Z_j(t), j = 1, 2, \cdots, 45$ as the heterogeneity induced by the nature of vowels and words together. Based on our observation from Figure 2, vowels (x-axis) and words (line symbols) demonstrate their own generic pattern, which are not necessarily additive; $U_{ijk}(t), k = 1, 2, \cdots, n_{ij}$ contain all the remaining variation within a specific speaker/word cell, including noise or measurement error. By applying the SFPCA algorithm, we extract the principal components as shown in Figure 7.

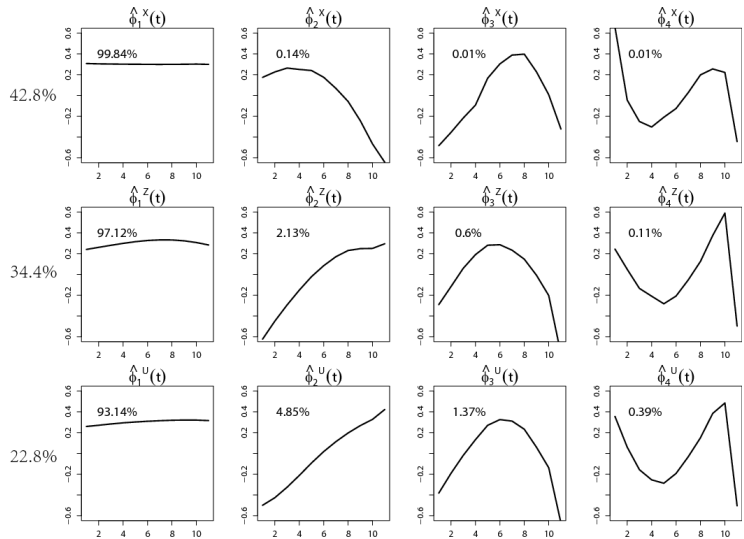

**Figure 7:** *Principal components for process $X_i$, $Z_j$ and $U_{ij}$ using two-way crossed model with sub-sampling (C2s). The first row show the first 4 principal components for the speaker-specific effect $X_i(t)$, while the second row display results for the word/vowel effect $Z_j(t)$. The proportion of variation explained by each principal components are listed in the plots. The estimated percentage of variation explained by each latent process is listed in front of the corresponding row.*

For the speaker-specific process, most of the variation (about 99.8%) of $X_i(t)$ is explained by the first PC which is essentially constant. Similarly, the first PC of the word/vowel-specific process $Z(t)$ remains constant. This finding is consistent with the original one of Aston *et al.* (2010): most of the variation arises from the mean 'shift' of the data. However, instead of modeling the overall PC scores to determine that the 'shift' is speaker- and word- dependent, we can claim that $\Phi_1^X(t)$ corresponds to speaker heterogeneity and $\Phi_1^Z(t)$ accounts for word/vowel difference. If we further care about the effects of speaker- or word-related covariates, we can fit regression models specifically to principal scores of each latent process. Also, the fact that the $2^{\text{nd}}$ - $4^{\text{th}}$ PCs of $Z_j(t)$ explain more variability than those of $X_i(t)$ indicates more complex structures induced by inherent vowel and word variability. Furthermore, with SFPCA, the relative effect size of speaker and word/vowel can be measured through the portion of variation in the whole dataset that each process explains (42.8% vs. 34.4% in Figure 7). Naturally, we are able to quantify the size of noise and measurement error through $U_{ij}(t)$. Note that we can easily characterize the reliability of speakers across the words spoken as $R_{XZ} = \text{trace}(\text{K}_\text{X} + \text{K}_\text{Z})/\text{trace}(\text{K}_\text{X} + \text{K}_\text{Z} + \text{K}_\text{U})$ (Shou *et al.* , 2012), which for this data set was 77.2%. The very interesting analysis provided by Aston *et al.* (2010) for this data analysis could not have shown these patterns or reliability estimates, as this would require an explicit modeling of the functional space. Instead, they focused on principal components that were common across processes and analyzed the scores using mixed effects models for scalar observations. Thus, their principal components are linear combinations of those shown in Figure 7, which may create confusion in the interpretation stage. The two approaches are complementary and should be pondered in particular applications.

## 5.2 Accelerometer data

In the accelerometer study, each participant has their activity intensity values recorded for 5 days during active periods (after waking up and before bedtime). Their active periods are identified using methods developed by Bai *et al.* (2012). Bai *et al.* (2012) mainly focuses on predicting movement type based on the three-axis accelerometer records. Here we are more interested in using the same dataset to assess the variability of energy expenditure in the population and from day to day. As Figure 1 indicates a periodic pattern every hour, we model the observed curves into three hierarchies: hours within days within each subject.

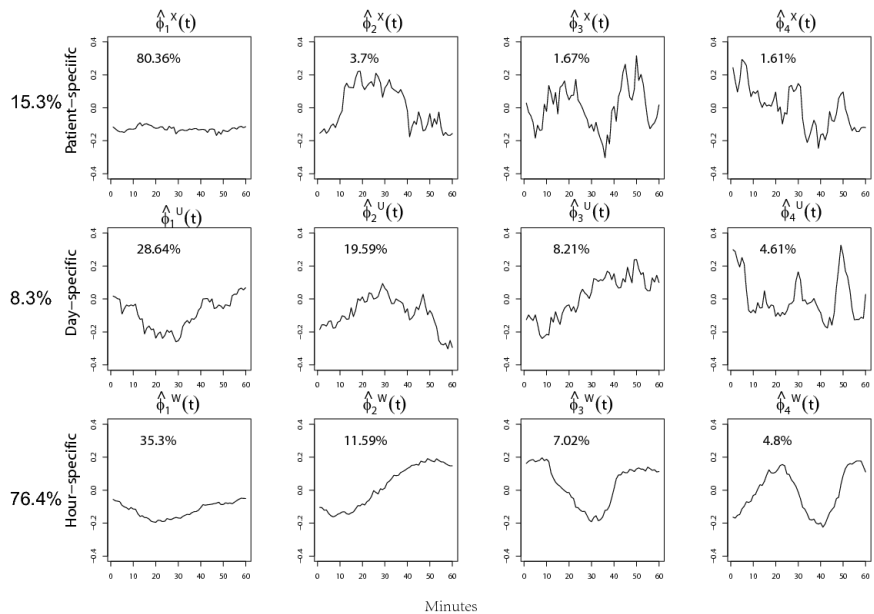

**Figure 8:** *Principal components for process $X$, $U$, and $W$ using three way nested model (N3). The proportion of variation explained by each PC component is listed in the plots. The first row show the first 4 PC components for the patient-specific effect $X(t)$, the second row display results for the day-specific effect $U(t)$ and the third row are estimated principal components for hour-specific effect $W(t)$. The proportion of variation explained by each latent process is labeled on the left side.*

The three-way nested model (N3) is applied to decompose the variance of the data. For the original dataset which contains 36,000 measurements per hour, we can implement SF-PCA using methods described in Section 3.3 for high-dimensional data. However, because data are densely sampled, it is more informative to smooth the data by averaging energy expenditure within every minute and conduct SFPCA on the summarized data. For simplicity, we also truncate the observations at the end of the study that does not complete an entire hour. Therefore, there are 60 measurements for every curve with a maximum of 19 curves per day for each subject. The first 4 principal components for the three levels of latent processes are displayed in Figure 8. The first component for the patient-specific process $X(t)$ accounts for the heterogeneity in the population. While the remaining few demonstrate either one-peak or double-peak energy expenditure pattern within one hour. Compared to subject-specific and hour-specific effects, the day-to-day variation (8.3%) accounts for a much smaller portion of the total variability. The majority (about 76%) of the total heterogeneity is contained in the hour-to-hour variation. This tells in a quantitative way that people follow a similar routine everyday, but their energy expenditure change dramatically within one day, depending on the type of activity they are involved in during a particular hour. The reliability ratio can also be evaluated as in the previous example.

# 6 Discussion

The defining characteristic of many functional studies is the existence of a specific structure in correlations vis-a-vis the experimental design, which can directly affect inference. Thus, there is an increasing demand for methods that 1) respect study design; 2) model multiple levels of variation; 3) are computationally feasible in high dimensions. In response to this demand, we have introduced a class of structured functional models that includes nested and crossed designs, and proposed a statistical framework, SFPCA, that estimates these models. Given the non-correlation assumption of latent processes, the covariance structure of the observed outcome is fully captured by the variance operators of the random processes. SFPCA is a set of efficient approaches that estimate and analyze the covariance structure using a uniform protocol for all the models. It uses functional PCA for dimensionality reduction and feature extraction.

The extensive simulation studies clearly demonstrated a great potential of the methodology to recover level-specific features of the latent processes. When we applied SFPCA to two studies that collected accelerometeric and phonetic data, we were able to distinguish various layers of effects inherent in the data. Similar to Section 5 in Koch (1967), our method can also be extended to the cases when the covariance matrix are heterogeneous and differ across levels.

Future work should focus on developing more efficient unbiased method of moment estimators that are adaptable to unbalanced designs. The development of combined methodology that infuses both 'naked' (nested/crossed) design-induced structures, with covariate-driven parts such as the one proposed in (Greven *et al.*, 2010), is an important, although challenging step in expanding this framework. Our methodology has a few potential limitations. Two most important ones are more rigorous treatment of noise (Di *et al.*, 2009) as well as possible sparsity in the functional observations Di *et al.* (2011).

# Acknowledgement

We thank Dr. John Aston for kindly providing us the phonetic study data and for his inspiring thoughts on the application of SFPCA.

# A  Finding principal scores for three-way nested and two-way crossed design

Here, we provide details of calculating principal scores in (N3) and (C2) models. We assume noise-free scenario and follow Zipunnikov *et al.* (2011). We can write model (N3) as $\widetilde{\mathbf{Y}}_i = \mathbf{B}\mathbf{u}_i$, where $\mathbf{B} = [\mathbf{B}_X|\mathbf{B}_U|\mathbf{B}_W]$ and $\mathbf{u}_i = (\boldsymbol{\xi}_i^{X'}, \boldsymbol{\xi}_i^{U'}, \boldsymbol{\xi}_i^{W'})'$. $\mathbf{B}_X = \mathbf{1}_{n_{i\cdot}} \otimes \boldsymbol{\Phi}_X$, $\mathbf{B}_U = \mathbf{I}_J \otimes (\mathbf{1}_{n_{ij\cdot}} \otimes \boldsymbol{\Phi}_U)$ and $\mathbf{B}_W = \mathbf{I}_{n_{i\cdot}} \otimes \boldsymbol{\Phi}_W$. So the BLUP of $\mathbf{u}_i$ would be

$$
\begin{aligned}
\hat{\mathbf{u}}_i &= (\mathbf{B}'\mathbf{B})^{-1}\mathbf{B}'\widetilde{\mathbf{Y}}_i \\
&= \begin{pmatrix} n_{i\cdot}\mathbf{I}_{N_1} & \mathbf{1}'_J \otimes (n_{ij}\boldsymbol{\Phi}'_X\boldsymbol{\Phi}_U) & \mathbf{1}'_{n_{i\cdot}} \otimes (\boldsymbol{\Phi}'_X\boldsymbol{\Phi}_W) \\ & \mathbf{I}_J \otimes (n_{ij}\mathbf{I}_{N_2}) & \mathbf{I}_J \otimes (\mathbf{1}'_{n_{ij}} \otimes \boldsymbol{\Phi}'_U\boldsymbol{\Phi}_W) \\ & & \mathbf{I}_{n_{i\cdot}N_3} \end{pmatrix}^{-1} \begin{pmatrix} \boldsymbol{\Phi}'_X\mathbf{Y}_i\mathbf{1}_{n_{i\cdot}} \\ \operatorname{vec}\{\boldsymbol{\Phi}'_U\mathbf{Y}_i(\mathbf{1}_{n_{ij}} \otimes \mathbf{I}_J)\} \\ \operatorname{vec}\{\boldsymbol{\Phi}'_W\mathbf{Y}_i\} \end{pmatrix} \\
&= \begin{pmatrix} n_{i\cdot}\mathbf{I}_{N_1} & \mathbf{1}'_J \otimes n_{ij}\mathbf{C}_{XU} & \mathbf{1}'_{n_{i\cdot}} \otimes \mathbf{C}_{XW} \\ & \mathbf{I}_J \otimes (n_{ij}\mathbf{I}_{N_2}) & \mathbf{I}_J \otimes (\mathbf{1}'_{n_{ij}} \otimes \mathbf{C}_{UW}) \\ & & \mathbf{I}_{n_{i\cdot}N_3} \end{pmatrix}^{-1} \begin{pmatrix} (\mathbf{A}_X^{N_1})'\mathbf{S}^{1/2}\mathbf{U}'_i\mathbf{1}_{n_{i\cdot}} \\ \operatorname{vec}\{(\mathbf{A}_U^{N_2})'\mathbf{S}^{1/2}\mathbf{U}'_i(\mathbf{1}_{n_{ij}} \otimes \mathbf{I}_J)\} \\ \operatorname{vec}\{(\mathbf{A}_W^{N_3})'\mathbf{S}^{1/2}\mathbf{U}'_i\} \end{pmatrix}
\end{aligned}
$$

where $\mathbf{C}_{XU} = (\mathbf{A}_X^{N_1})'\mathbf{A}_U^{N_2}$ and $\mathbf{C}_{XW}$, $\mathbf{C}_{UW}$ are similarly defined.

For two-way crossed model (C2) with balanced design, $\widetilde{\mathbf{Y}} = \mathbf{B}\mathbf{u}$, where $\widetilde{\mathbf{Y}} = (\mathbf{Y}'_{11}, \cdots, \mathbf{Y}'_{IJ})'$. $\mathbf{B} = [\mathbf{B}_X|\mathbf{B}_Z|\mathbf{B}_W]$, where $\mathbf{B}_X = \mathbf{I}_I \otimes (\mathbf{1}_J \otimes \boldsymbol{\Phi}_X)$, $\mathbf{B}_Z = \mathbf{1}_I \otimes (\mathbf{I}_J \otimes \boldsymbol{\Phi}_Z)$ and $\mathbf{B}_W = \mathbf{I}_{IJ} \otimes \boldsymbol{\Phi}_W$; $\mathbf{u} = (\mathbf{u}^X|\mathbf{u}^Z|\mathbf{u}^W)'$ and $\mathbf{u}^X = (\boldsymbol{\xi}_1^{X'}, \cdots, \boldsymbol{\xi}_I^{X'})$, $\mathbf{u}^Z = (\boldsymbol{\xi}_1^{Z'}, \cdots, \boldsymbol{\xi}_J^{Z'})$ and $\mathbf{u}^W = (\boldsymbol{\xi}_{11}^{W'}, \cdots, \boldsymbol{\xi}_{IJ}^{W'})^T$. The BLUP gives

$$
\begin{aligned}
\hat{\mathbf{u}} &= (\mathbf{B}'\mathbf{B})^{-1}\mathbf{B}'\widetilde{\mathbf{Y}} \\
&= \begin{pmatrix} J\mathbf{I}_{IN_1} & (\mathbf{1}_I\mathbf{1}'_J)\otimes(\mathbf{\Phi}'_X\mathbf{\Phi}_Z) & (\mathbf{I}_I\otimes\mathbf{1}'_J)\otimes(\mathbf{\Phi}'_X\mathbf{\Phi}_W) \\ & I\mathbf{I}_{JN_2} & (\mathbf{1}'_I\otimes\mathbf{I}_J)\otimes(\mathbf{\Phi}'_Z\mathbf{\Phi}_W) \\ & & \mathbf{I}_{IJN_3} \end{pmatrix}^{-1} \begin{pmatrix} \mathrm{vec}\{\mathbf{\Phi}'_X\mathbf{Y}(\mathbf{I}_I\otimes\mathbf{1}_J)\} \\ \mathrm{vec}\{\mathbf{\Phi}'_Z\mathbf{Y}(\mathbf{1}_I\otimes\mathbf{I}_J)\} \\ \mathrm{vec}(\mathbf{\Phi}'_W\mathbf{Y}) \end{pmatrix} \\
&= \begin{pmatrix} J\mathbf{I}_{IN_1} & (\mathbf{1}_I\mathbf{1}'_J)\otimes\mathbf{C}_{XZ} & (\mathbf{I}_I\otimes\mathbf{1}'_J)\otimes\mathbf{C}_{XW} \\ & I\mathbf{I}_{JN_2} & (\mathbf{1}'_I\otimes\mathbf{I}_J)\otimes\mathbf{C}_{ZW} \\ & & \mathbf{I}_{IJN_3} \end{pmatrix}^{-1} \begin{pmatrix} \mathrm{vec}\{(\mathbf{A}_X^{N_1})'\mathbf{S}^{1/2}\mathbf{U}'(\mathbf{I}_I\otimes\mathbf{1}_J)\} \\ \mathrm{vec}\{(\mathbf{A}_Z^{N_2})'\mathbf{S}^{1/2}\mathbf{U}'(\mathbf{1}_I\otimes\mathbf{I}_J)\} \\ \mathrm{vec}\{(\mathbf{A}_W^{N_3})'\mathbf{S}^{1/2}\mathbf{U}'\} \end{pmatrix}
\end{aligned}
$$

Again, $\mathbf{C}_{XZ} = \mathbf{\Phi}'_X\mathbf{\Phi}_Z = (\mathbf{A}_X^{N_1})'\mathbf{A}_Z^{N_2}$, same for $\mathbf{C}_{XW}$ and $\mathbf{C}_{ZW}$.

# B MoM estimators for additional models

## B.1 Multi-way nested model (NM)

The covariance operators of latent processes in multi-way nested model satisfy

$$
E\left[\{Y_{i_1i_2\cdots i_r}(t) - Y_{h_1h_2\cdots h_r}(t)\}\{Y_{i_1i_2\cdots i_r}(s) - Y_{h_1h_2\cdots h_r}(s)\}^T\right]
$$

$$
\begin{aligned}
&= 2K_r(t,s), && \text{if } i_1 = h_1, \cdots, i_{r-1} = h_{r-1}, i_r \neq h_r \\
&= 2\{K_{r-1}(t,s) + K_r(t,s)\}, && \text{if } i_1 = h_1, \cdots, i_{r-2} = h_{r-2}, i_{r-1} \neq h_{r-1} \\
&\cdots \\
&= 2\{K_1(t,s) + \cdots + K_r(t,s)\}, && \text{if } i_1 \neq h_1
\end{aligned}
$$

Therefore, we have $H_j(j=1,2,\cdots,r)$ operators as

$$
\begin{aligned}
H_1 &= \frac{1}{k_1 - n} \sum_{i_1,i_2,\cdots,i_{r-1}} \sum_{i_r,h_r} \{Y_{i_1\cdots i_{r-1}i_r} - Y_{i_1\cdots i_{r-1}h_r}\}\{Y_{i_1\cdots i_{r-1}i_r} - Y_{i_1\cdots i_{r-1}h_r}\}^T = \frac{2}{k_1 - n}\mathbf{Y}(\mathbf{D}_1 - \mathbf{E}_1^T\mathbf{E}_1)\mathbf{Y}^T \\
H_2 &= \frac{1}{k_2 - k_1} \sum_{i_1,\cdots,i_{r-2}} \sum_{i_{r-1}\neq h_{r-1}} \sum_{i_r,h_r} (Y_{i_1\cdots i_{r-2}i_{r-1}i_r} - Y_{i_1\cdots i_{r-2}h_{r-1}h_r})(Y_{i_1\cdots i_{r-2}i_{r-1}i_r} - Y_{i_1\cdots i_{r-2}h_{r-1}h_r})^T \\
&= \frac{2}{k_2 - k_1}\mathbf{Y}(\mathbf{D}_2 - \mathbf{E}_2^T\mathbf{E}_2 - \mathbf{D}_1 + \mathbf{E}_1^T\mathbf{E}_1)\mathbf{Y}^T \\
H_3 &= \frac{1}{k_3 - k_2}\Big\{ \sum_{\substack{i_{r-1},i_r,h_{r-1},h_r \\ i_{r-2}\neq h_{r-2} \\ i_1,\cdots,i_{r-3}}} (Y_{i_1\cdots i_{r-3}i_{r-2}i_{r-1}i_r} - Y_{i_1\cdots i_{r-3}h_{r-2}h_{r-1}h_r})(Y_{i_1\cdots i_{r-3}i_{r-2}i_{r-1}i_r} - Y_{i_1\cdots i_{r-3}h_{r-2}h_{r-1}h_r})^T \\
&\qquad - \sum_{\substack{i_1,\cdots,i_{r-2} \\ i_{r-1},i_r,h_{r-1},h_r}} (Y_{i_1\cdots i_{r-2}i_{r-1}i_r} - Y_{i_1\cdots i_{r-2}h_{r-1}h_r})(Y_{i_1\cdots i_{r-2}i_{r-1}i_r} - Y_{i_1\cdots i_{r-2}h_{r-1}h_r})^T \Big\} \\
&= \frac{2}{k_3 - k_2}\mathbf{Y}(\mathbf{D}_3 - \mathbf{E}_3^T\mathbf{E}_3 - \mathbf{D}_2 + \mathbf{E}_2^T\mathbf{E}_2)\mathbf{Y}^T \\
&\cdots \\
H_j &= \frac{2}{k_j - k_{j-1}}\mathbf{Y}(\mathbf{D}_j - \mathbf{E}_j^T\mathbf{E}_j - \mathbf{D}_{j-1} + \mathbf{E}_{j-1}^T\mathbf{E}_{j-1})\mathbf{Y}^T \\
&\cdots
\end{aligned}
$$

where $k_j = \sum_{i_1 i_2 \cdots i_{r-j}} n_{i_1 i_2 \cdots i_{r-j}}^2$, $j = 1, 2, \cdots, r$. The covariance operators are represented as

$$
\begin{aligned}
\widehat{K}_{r+1-j} &= (H_{j+1} - H_j)/2 \\
&= \mathbf{Y}\Big\{\frac{1}{k_{j+1} - k_j}(\mathbf{D}_{j+1} - \mathbf{E}_{j+1}^T\mathbf{E}_{j+1} - \mathbf{D}_j + \mathbf{E}_j^T\mathbf{E}_j) - \frac{1}{k_j - k_{j-1}}(\mathbf{D}_j - \mathbf{E}_j^T\mathbf{E}_j - \mathbf{D}_{j-1} + \mathbf{E}_{j-1}^T\mathbf{E}_{j-1})\Big\}\mathbf{Y}^T
\end{aligned}
$$

## B.2 Two-way crossed design with sub-sampling (C2s)

With model $Y_{ijk}(t) = \mu(t) + X_i(t) + Z_j(t) + W_{ij}(t) + U_{ijk}(t)$, $i = 1, 2, \cdots, I$; $j = 1, 2, \cdots, J$ and $k = 1, 2, \cdots, n_{ij}$, we have

$$
\begin{aligned}
E\{Y_{ijk}(t) - Y_{luv}(t)\}&\{(Y_{ijk}(s) - Y_{luv}(s)\}^T \\
&= 2K_U(t,s), &&\text{if } i = l, j = u, k \neq v \\
&= 2\{K_Z(t,s) + K_W(t,s) + K_U(t,s)\}, &&\text{if } i = l, j \neq u \\
&= 2\{K_X(t,s) + K_W(t,s) + K_U(t,s)\}, &&\text{if } i \neq l, j = u \\
&= 2\{K_X(t,s) + K_Z(t,s) + K_W(t,s) + K_U(t,s)\}, &&\text{if } i \neq k, j \neq u
\end{aligned}
$$

The corresponding $H$ operators are

$$
\begin{aligned}
H_U &= \frac{1}{k_{12}-n}\sum_{i,j}\sum_{k\neq v}(Y_{ijk}-Y_{ijv})(Y_{ijk}-Y_{ijv})^T = \frac{2}{k_{12}-n}\mathbf{Y}(\mathbf{D}_{12}-\mathbf{E}_{12}^T\mathbf{E}_{12})\mathbf{Y}^T \\
H_Z &= \frac{1}{k_1-k_{12}}\sum_{i=1}\sum_{j\neq u}\sum_{k,v}(Y_{ijk}-Y_{iuv})(Y_{ijk}-Y_{iuv})^T = \frac{2}{k_1-k_{12}}\mathbf{Y}(\mathbf{D}_1-\mathbf{E}_1^T\mathbf{E}_1-\mathbf{D}_{12}+\mathbf{E}_{12}^T\mathbf{E}_{12})\mathbf{Y}^T \\
H_X &= \frac{1}{k_2-k_{12}}\sum_{i\neq l}\sum_{j}\sum_{k,v}(Y_{ijk}-Y_{ljv})(Y_{ijk}-Y_{ljv})^T = \frac{2}{k_2-k_{12}}\mathbf{Y}(\mathbf{D}_2-\mathbf{E}_2^T\mathbf{E}_2-\mathbf{D}_{12}+\mathbf{E}_{12}^T\mathbf{E}_{12})\mathbf{Y}^T \\
H_T &= \frac{1}{n^2-k_1-k_2+k_{12}}\sum_{i\neq l}\sum_{j\neq u}\sum_{k,v}(Y_{ijk}-Y_{luv})(Y_{ijk}-Y_{luv})^T \\
&= \frac{2}{n^2-k_1-k_2+k_{12}}\mathbf{Y}(n\mathbf{I}-\mathbf{1}\mathbf{1}^T-\mathbf{D}_1-\mathbf{D}_2+\mathbf{D}_{12}+\mathbf{E}_1^T\mathbf{E}_1+\mathbf{E}_2\mathbf{E}_2-\mathbf{E}_{12}\mathbf{E}_{12}^T)\mathbf{Y}^T
\end{aligned}
$$

where $k_1 = \sum_i n_{i0}^2$, $k_2 = \sum_j n_{0j}^2$, $k_{12} = \sum_{i,j} n_{ij}^2$

## B.3 Multi-way crossed design (CM)

The most general model for crossed design is

$$Y_{i_1 i_2 \cdots i_r u}(t) = \mu(t) + R_{i_1}^{(1)}(t) + R_{i_2}^{(2)}(t) + \cdots + R_{i_r}^{(r)}(t) + \cdots + R_{i_{j_1} i_{j_2} \cdots i_{j_q}}^{(j_1 j_2 \cdots j_q)}(t) + \cdots + R_{i_1 i_2 \cdots i_r}^{(12\cdots r)}(t) + R_{i_1 i_2 \cdots i_r u}(t)$$

with $i_k = 1, 2, \cdots, m_k$ where $k = 1, 2, \cdots, r$, $u = 1, 2, \cdots, n_{i_1 i_2 \cdots i_r}$; $1 \leq q \leq r$, $(j_1, j_2, \cdots, j_q) \in \{1, 2, \cdots, r\}$, $j_1 < j_2 < \cdots < j_q$ and $R_{i_{j_1} i_{j_2} \cdots i_{j_q}}^{(j_1 j_2 \cdots j_q)}(t)$ has variance operator $K_{j_1 j_2 \cdots j_q}$, $R_{i_1 i_2 \cdots i_r u}(t)$ has covariance operator $K_U$. With similar procedure as in the previous sections, we can work out the formula for the covariance operators. We omit the details here.

## B.4 Experiments of the Mixed Type

Given $Y_{ijkl}(t) = \mu(t) + X_i(t) + Z_j(t) + W_{ij}(t) + C_{ik}(t) + D_{jl}(t) + U_{ijk}(t) + V_{ijl}(t) + E_{ijkl}(t)$,

$$E\{(Y_{ijkl}(t) - Y_{i'j'k'l'}(t))(Y_{ijkl}(s) - Y_{i'j'k'l'}(s))^T\}$$

$$= 2\{K_D(t,s) + K_V(t,s) + K_E(t,s)\}, \text{ if } i=i', j=j', k=k', l \neq l'$$

$$= 2\{K_C(t,s) + K_U(t,s) + K_E(t,s)\}, \text{ if } i=i', j=j', k \neq k', l = l'$$

$$= 2\{K_C(t,s) + K_D(t,s) + K_U(t,s) + K_V(t,s) + K_E(t,s)\}, \text{ if } i=i', j=j', k \neq k', l \neq l'$$

$$= 2\{K_Z(t,s) + K_W(t,s) + K_D(t,s) + K_U(t,s) + K_V(t,s) + K_E(t,s)\}, \text{ if } i=i', j \neq j', k = k'$$

$$= 2\{K_Z(t,s) + K_W(t,s) + K_C(t,s) + K_D(t,s) + K_U(t,s) + K_V(t,s) + K_E(t,s)\}, \text{ if } i=i', j \neq j', k \neq k'$$

$$= 2\{K_X(t,s) + K_W(t,s) + K_C(t,s) + K_U(t,s) + K_V(t,s) + K_E(t,s)\}, \text{ if } i \neq i', j = j', l = l'$$

$$= 2\{K_X(t,s) + K_W(t,s) + K_C(t,s) + K_D(t,s) + K_U(t,s) + K_V(t,s) + K_E(t,s)\}, \text{ if } i \neq i', j = j', l \neq l'$$

$$= 2\{K_X(t,s) + K_Z(t,s) + K_W(t,s) + K_C(t,s) + K_D(t,s) + K_U(t,s) + K_V(t,s) + K_E(t,s)\}, \text{ if } i \neq i', j \neq j'$$

So the $H$ operators are expressed as

$$H_L = \frac{1}{\sum_{i,j,k} n_{ijk0}^2 - \sum_{i,j,l,k} n_{ijkl}^2} \sum_{i,j,k} \sum_{l \neq l'} (Y_{ijkl} - Y_{ijkl'})(Y_{ijkl} - Y_{ijkl'})^T$$

$$H_K = \frac{1}{\sum_{i,j,l} n_{ij0l}^2 - \sum_{i,j,k,l} n_{ijkl}^2} \sum_{i,j,l} \sum_{k \neq k'} (Y_{ijkl} - Y_{ijk'l})(Y_{ijkl} - Y_{ijk'l})^T$$

$$H_E = \frac{1}{\sum_{i,j} n_{ij00}^2 + 2\sum_{i,j,l,k} n_{ijkl}^2 - \sum_{i,j,k} n_{ijk0}^2 - \sum_{i,j,l} n_{ij0l}^2} \sum_{i,j} \sum_{l \neq l', k \neq k'} (Y_{ijkl} - Y_{ijk'l'})(Y_{ijkl} - Y_{ijk'l'})^T$$

$$H_J = \frac{1}{\sum_{i,k} n_{i0k0}^2 - \sum_{i,j,k} n_{ijk0}^2} \sum_{i} \sum_{j \neq j'} \sum_{l,l',k} (Y_{ijkl} - Y_{ij'kl'})(Y_{ijkl} - Y_{ij'kl'})^T$$

$$H_{JK} = \frac{1}{\sum_{i} n_{i000}^2 - \sum_{i,j} n_{ij00}^2 - \sum_{i,k} n_{i0k0}^2 + \sum_{i,j,k} n_{ijk0}^2} \sum_{i} \sum_{j \neq j'} \sum_{l,l',k \neq k'} (Y_{ijkl} - Y_{ij'k'l'})(Y_{ijkl} - Y_{ij'k'l'})^T$$

$$H_I = \frac{1}{\sum_{j,l} n_{0j0l}^2 - \sum_{i,j,l} n_{ij0l}^2} \sum_{j} \sum_{i \neq i'} \sum_{l,k',k} (Y_{ijkl} - Y_{i'jk'l})(Y_{ijkl} - Y_{i'jk'l})^T$$

$$H_{IL} = \frac{1}{\sum_{j} n_{0j00}^2 - \sum_{i,j} n_{ij00}^2 - \sum_{j,l} n_{0j0l}^2 + \sum_{i,j,l} n_{ij0l}^2} \sum_{i \neq i'} \sum_{j} \sum_{k',k,l \neq l'} (Y_{ijkl} - Y_{i'jk'l'})(Y_{ijkl} - Y_{i'jk'l'})^T$$

$$H_{IJ} = \frac{1}{n^2 + 2\sum_{i,j} n_{ij00}^2 - \sum_{i} n_{i000}^2 - \sum_{j} n_{0j00}^2} \sum_{i \neq i', j \neq j'} \sum_{k',k,l,l'} (Y_{ijkl} - Y_{i'j'k'l'})(Y_{ijkl} - Y_{i'j'k'l'})^T$$

# C  Additional simulation results

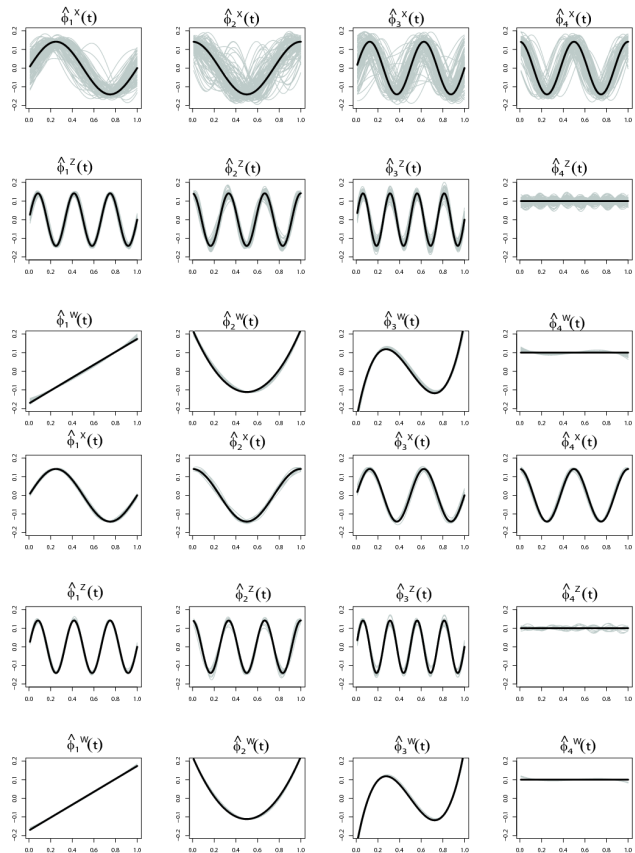

**Figure 9:** *The first 3 panels are estimated principal components when there are 20 samples in $X(t)$ and 200 in $Z(t)$. The bottom 3 panels are results from the same model as in equation (2) when $I = J = 200$.*

# References

Aston, J. A. D., Chiou, J. M., & Evans, J. P. 2010. Linguistic pitch analysis using functional principal component mixed effect models. *Jounral of the Royal Statistical Society, Series C*, **59**(2), 297–317.

Bai, J., Goldsmith, J., Caffo, B. S., Glass, T., & Crainiceanu, C. M. 2012. Movelets : a dictionary of movement. *Electronic Journal of Statistics*, **6**, 559–578.

Brumback, B. A., & Rice, J. A. 1998. Smoothing spline models for the analysis of nested and crossed samples of curves. *Journal of the American Statistical Association*, **93**(443), 961–976.

Crainiceanu, C. M., Caffo, B. S., Di, C. Z., & Punjabi, N. M. 2009. Nonparametric signal extraction and measurement error in the analysis of electroencephalographic activity during sleep. *Journal of American Statistical Association*, **104**(486), 541–555.

Di, C. Z., Crainiceanu, C. M., Caffo, B. S., & Punjabi, N. M. 2009. Multilevel functional principal component analysis. *The Annals of Applied Statistics*, **3**(1), 458–488.

Di, C. Z., Jank, W. S., & Crainiceanu, C. M. 2011. Multilevel sparse functional principal component analysis. *Under invited revisions for Computational Statistics and Data Analysis*.

Greven, S., Crainiceanu, C. M., Caffo, B. S., & Reich, D. 2010. Longitudinal functional principal component analysis. *Electronic journal of statistics*, **4**, 1022–1054.

Guo, W. 2002. Functional mixed effects models. *Biometrics*, **58**(1), 121–128.

Guo, W. 2004. Functional data analysis in longitudinal settings using smoothing splines. *Statistical methods in medical research*, **13**(1), 49–62.

Herrick, R. C., & Morris, J. S. 2006. Wavelet-based functional mixed model analysis: computation considerations. *In Proceedings, Joint Statistical Meetings, ASA Section on Statistical Computing*.

Karhunen, K. 1947. Über lineare Methoden in der Wahrscheinlichkeitsrechnung. *Ann. Acad. Sci. Fennicae. Ser. A. I. Math.-Phys.*, **37**, 1–79.

Koch, G. G. 1967. A general approach to the estimation of variance components. *Technometrics*, **9**(1), 93–118.

Koch, G. G. 1968. Some further remarks concerning "A general approach to the estimation of variance components". *Technometrics*, **10**(3), 551–558.

Loève, M. 1978. *Probability theory*. 4th edn. Springer-Verlag. ISBN 0-387-90262-7.

Morris, J. S., & Carroll, R. J. 2006. Wavelet-based functional mixed models. *Journal of the Royal Statistical Society, Series B*, **68**(2), 179–199.

Quan, S. F., Howard, B. V., Iber, C., Kiley, J. P., Nieto, F. J., OConnor, G. T., Rapoport, D. M., Redline, S., Robbins, J., Samet, J. M., & Wahl, P. W. 1997. The sleep heart health study: design, rationale, and methods. *Sleep*, **20**(12), 1077– 1085.

Ramsay, J. O., & Silverman, B. 2005. *Functional data analysis*. Second edn. Springer. MR2168993.

Shou, H., Eloyan, A., Lee, S., Zipunnikov, V., Caffo, B. S., Lindquist, M., & Crainiceanu, C. M. 2012. The image intra-class correlation coefficient (I2C2) for replication studies. *Under revision for Cognitive, Affective, and Behavioral Neuroscience*.

Staicu, A. M., Crainiceanu, C. M., & Carroll, R. J. 2010. Fast methods for spatially correlated multilevel functional data. *Biostatistics*, **11**(2), 177–194.

Staniswalis, J. G., & Lee, J. J. 1998. Nonparametric Regression Analysis of Longitudinal Data. *Journal of the American Statistical Association*, **93**(444), 1403–1418.

Yao, F., Clifford, A.J., Dueker, S.R., Follett, J., Lin, Y., Buchholz, B.A., & Vogel, J.S. 2003. Shrinkage estimation for functional principal component scores with application to the population kinetics of plasma folate. *Biometrics*, **59**(3), 676–685.

Yao, F., Müller, H.G., & Wang, J.L. 2005. Functional data analysis for sparse longitudinal data. *Journal of the American Statistical Association*, **100**(470), 577–590.

Zhou, L., Huang, J. Z., Martinez, J. G., Maity, A., Baladandayuthapani, V., & Carroll, R. J. 2010. Reduced rank mixed effects models for spatially correlated hierarchical functional data. *Journal of the American Statistical Association*, **105**(489), 390–400.

Zipunnikov, V., Caffo, B. S., Yousem, D. M., Davatzikos, C., Schwartz, B. S., & Crainiceanu, C. M. 2011. Multilevel functional principal component analysis for high-dimensional data. *Journal of Computational and Graphical Statistics*, **20**(4), 852–873.